**On the project risk baseline: integrating aleatory uncertainty into project scheduling**

**Abstract**

Obtaining a viable schedule baseline that meets all project constraints is one of the main issues for project managers. The literature on this topic focuses mainly on methods to obtain schedules that meet resource restrictions and, more recently, financial limitations. The methods provide different viable schedules for the same project, and the solutions with the shortest duration are considered the best-known schedule for that project. However, no tools currently select which schedule best performs in project risk terms. To bridge this gap, this paper aims to propose a method for selecting the project schedule with the highest probability of meeting the deadline of several alternative schedules with the same duration. To do so, we propose integrating aleatory uncertainty into project scheduling by quantifying the risk of several execution alternatives for the same project. The proposed method, tested with a well-known repository for schedule benchmarking, can be applied to any project type to help managers to select the project schedules from several alternatives with the same duration, but the lowest risk.

**Keywords**: Total Project Risk; Aleatory Uncertainty; Schedule Risk Value; Schedule Risk Baseline; Project Risk Management.

## 1. Introduction

Obtaining a project schedule is a complex process whose output is conditioned by many decisions made by project managers. Project scheduling is considered the last step of the initial project planning cycle (Pellerin and Perrier, 2019). This schedule must comply with all the project constraints identified by project managers and is later used as a baseline for project execution (Afshar-Nadjafi, 2016).

Following PMBOK's guide (Project Management Institute, 2017), the basis for obtaining a project schedule is the project activity list, which derives from the work packages described in the work breakdown structure (WBS) containing all the work defined to complete the project's scope. Based on this activity list, and following a series of restrictions (durations and precedence relationships), project managers may obtain a project schedule by following classic scheduling techniques like the Critical Path Method (CPM) (Kelley and Walker, 1959) or the Program Evaluation and Review Technique (PERT) (Malcolm et al., 1959), which deliver a schedule in which each project activity is set to start at its earliest start time (i.e. activities are scheduled to start as soon as possible based on precedence relationships). This schedule is often referred to as the 'critical path schedule' and provides the shortest possible project duration (Fig. 1., a).

For this preliminary schedule to become the schedule baseline (i.e. the reference to implement project execution), project managers must ensure that it meets other constraints that arise from project planning (i.e. not only deadline constraints, but also the contractual objectives of costs, quality, and satisfaction of stakeholders). Although PMBOK proposes a succession of processes in each knowledge area, obtaining a schedule baseline is a complex goal given the complex interdependence among processes (please refer to Ruiz-Martin and Poza (2015) for a detailed procedure that guarantees coherence among knowledge areas in a project management plan). As far as project scheduling is concerned, obtaining a feasible schedule requires considering mainly the availability of both resources and funds.

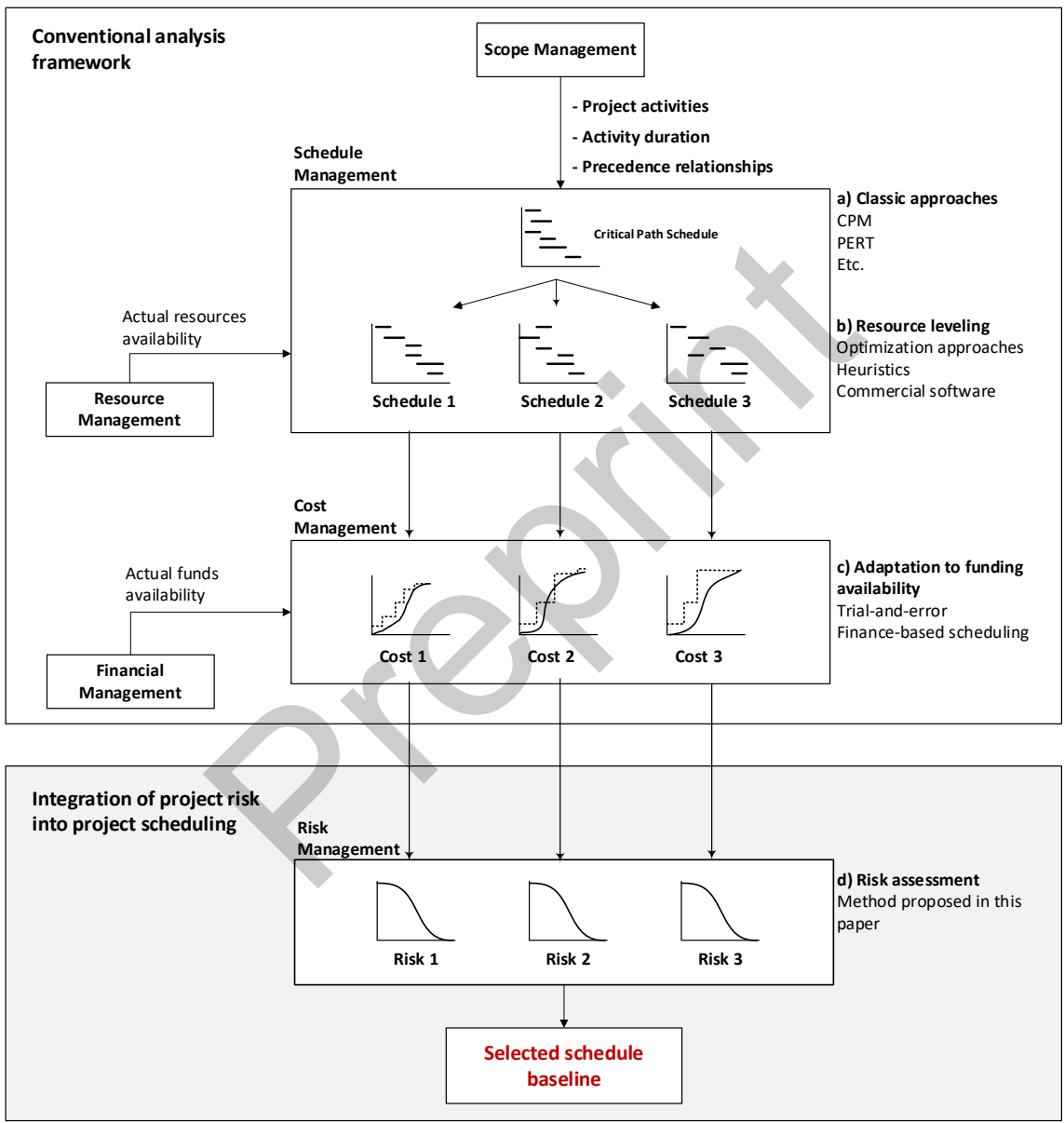

**Fig. 1.** A simplified scheme of the steps to obtain a feasible schedule for a project

When it comes to resource constraints, the literature has traditionally focused on resource leveling (i.e. obtaining a schedule that does not exceed the number of resources available per time unit). Resource leveling requires delaying tasks until resources become available. This problem has been widely studied in the Operations Research field and is named resource-constrained project scheduling problem (RCPSP) (Fendley, 1968). However, obtaining such a schedule entails a complex combinatorial problem for which there is no optimal solution (Blazewicz et al., 1983; Villafáñez et al., 2018). For this reason, proposals for solving the RCPSP abound in the literature, which may be based on optimization methods (Karam and Lazarova-Molnar, 2013) or heuristics (Pellerin et al., 2020; Villafáñez et al., 2019). To this end, project management software, such as Microsoft Project or Oracle Primavera, is widely used by professionals (Dasović et al., 2020; Hazir, 2015). However, as the optimal solution is unknown, the application of any of these techniques provides different schedules for the same project (Fig. 1., b). This means that, although these schedules comply with resource constraints (i.e. there is no resource overallocation), there is no guarantee that the obtained schedule corresponds to the optimal solution (and is subject to project managers' criteria).

Although the schedules obtained by the above methods are feasible insofar that resources availability is guaranteed, they do not ensure funds availability to execute project activities. A leveled schedule, along with activity cost estimates, allows a project cost baseline to be obtained (i.e. the cumulative estimated cost during its life cycle). At this point, the cost baseline associated with each schedule may be compared to the actual funding available for project execution (Fig. 1., c). If project managers detect that costs exceed actual funding at some point, they may delay some activities to ensure funds availability. This process can be done manually on a trial-and-error basis. However, Elazouni and Gab-Allah (2004) introduced the so-called project-based scheduling, which considers both resources availability and the activities cost in the scheduling process to ensure a workable schedule in terms of both resource leveling and funding requirements. Since then, other methods along these lines have been proposed (Alavipour and Arditi, 2019; Pinha and Ahluwalia, 2019; Villafáñez et al., 2020).

As indicated above, there might be many feasible schedules to implement the same project scope. Therefore, project managers need tools to select the schedule that best matches the company's needs. Traditionally, the scope baseline, the schedule baseline and the cost baseline have been considered the triad to make these decisions (Taylor, 2008).

In this paper, however, we introduce the risk baseline as an additional decision tool for project managers. The proposed method intends to assist project managers in selecting the best feasible schedule for the project by considering variability (in the form of aleatory uncertainty) in the duration of each schedule. The starting point of the proposed method is the set of schedules that comply with both resource allocation and available funding. In all these schedules, the total risk is quantified (Fig. 1., d) so that project managers are able to select a schedule baseline based not only on deadline or funding terms, but also on the total risk associated with each schedule.

The rest of the paper is structured as follows. Section 2 reviews the project risk and uncertainty concepts as used by the scientific community, and specifies the sense in which these terms are herein used: Among the different types of uncertainty that can impact the total project risk (i.e. aleatory, epistemic, stochastic and ontological), this paper focuses on aleatory uncertainty. Section 3 presents the tools and indicators to measure and quantify the total project risk on which our approach is based. Section 4 illustrates how to apply the proposed procedure with an example. Section 5 presents and discusses the results obtained after applying our approach to demonstrate that it can be used to select the schedule with the lowest risk among all the feasible alternatives with the same duration. Finally, Section 6 offers the conclusions drawn from this work.

## 2. Literature review

All projects are risky because, by definition, they result from performing a unique activity with some degree of complexity and uncertainty (Deshmukh et al., 2020; Dey et al., 2013; El-Sayegh et al., 2018; Farooq et al., 2018; Hillson, 2009; Kimiagari and Keivanpour, 2019, etc). However, a literature review shows that there has not always been a consensus about the 'risk' concept (Williams, 1995).

At first, the authors attributed only negative connotations to its meaning insofar as a risk always resulted in adverse outcomes for the project (Dowie, 1999). However, the risk concept evolved to also include positive aspects (i.e. the so-called opportunities) (Chapman and Ward, 2003; Hillson, 2002a; Hillson and Simon, 2012; Jaafari, 2001). This extension of the risk concept was incorporated by practitioners and academics (Hillson, 2002b), and also by the main risk management standards (Association for Project Management, 2004; European Commission, 2018; International Standards Organisation, 2018; OGC, 2009; Project Management Institute, 2017, 2009).

Hillson (2009) defines risk as "an uncertainty that, if it occurred, could affect one objective or more". The author considers that there are many uncertainties, but only those that can affect the project can be considered risks. In other words, according to this definition, risk is understood as "uncertainty that matters". More recent works also contemplate this concept whereby risk emerges from uncertainty (Hillson and Simon, 2012) or the outcome of uncertainty on objectives (Alleman et al., 2018).

According to this view, the 'risk' concept is related to the 'uncertainty' concept. Similarly to that observed with the 'risk' concept, different authors also employ the term 'uncertainty' to refer to different concepts. Perminova et al. (2008) defines uncertainty as "an event or a situation, which was not expected to happen, regardless of whether it could have been possible to consider it in advance". Alleman et al. (2018) defines it as a "state or condition that involves a deficiency of information and leads to inadequate or incomplete knowledge or understanding". While some authors use the term uncertainty in a general sense (i.e. referring to lack of certainty) (Howell et al., 2010), others relate uncertainty to the objectives or methods used in the project (Crawford et al., 2006; Millington and Stapleton, 2005; Pearson, 1990; Turner and Cochrane, 1993) with: the market or technology (Jordan et al., 2005; Shenhar and Dvir, 2007); changes in the project (Little, 2005); external influences (Ratbe et al., 1999).

Beyond the different possible uncertainty definitions, several works (Elms, 2004; Frank, 1999; Schafer, 1976) argue the need to distinguish between two different uncertainty types: *aleatory uncertainty* (which is embedded practically in each activity, e.g. range of duration for many reasons) and *epistemic uncertainty* (due to ambiguity or imperfect knowledge). Therefore by extending this uncertainty conception, Hillson (2014c) argues that two additional uncertainty types should be added to the above classification: *stochastic uncertainty* (also called 'event risk', defined as 'future possible events') and *ontological uncertainty* (also called 'unknown-unknowns', which is unknown knowledge of what is impossible to know).

Returning to the risk concept in project management, later works distinguish between 'individual risks' and the 'overall project risk'. Individual risks affect one activity or more in the same project, whereas the overall project risk is defined as "the effect of uncertainty on the project as a whole" (Hillson, 2014b, 2014c). The overall project risk is "more than the sum of individual risks within a project because it includes all sources of project uncertainty". As individual risks and overall project risks affect the project at distinct levels, radically different approaches are required to manage them (Hillson, 2014c).

The literature review shows that there is no consensus about the risk and uncertainty concepts. In this context, we herein use the risk concept in line with Hillson (2009). In addition, of the four uncertainty types suggested by Hillson (2014c), we focus on aleatory uncertainty.

Project managers are quite clear about how to assess each individual risk, mainly through qualitative analyses in which probability-impact matrices allow a value to be assigned to each identified risk (Chapman and Ward, 2000; Cox, 2008; El-Sayegh et al., 2018; Emblemsvåg and Kjølstad, 2006; Fergany et al., 2020; Ward, 1999). This type of analysis assigns a each risk certain degree of relevance, which allows the identified risks to be prioritized according to their relevance (Chapman, 2006; Chapman and Ward, 2003; Project Management Institute, 2017).

However, the project risk on the whole (i.e., overall project risk, hereafter the project risk) cannot be understood only as the sum of the identified individual risks. Some authors use fuzzy techniques to measure the project risk (Doskočil, 2015; Gavrysh and Melnykova, 2019; Ghaffari et al., 2014; Hsieh et al., 2018; Liu et al., 2007; Xie et al., 2006). Quantitative techniques, such as Monte Carlo simulation, have been employed extensively to estimate the project risk (Hulett, 2011; Vose, 2008). Monte Carlo simulation is adequate for this quantitative analysis because it presents a range of possible results, as well as the probability of these results being achieved (Acebes et al., 2015, 2014a; Khedr, 2006; Kwak and Ingall, 2007; Rezaie et al., 2007; Wirawan and Garniwa, 2018). The possible project outcomes are usually shown as a distribution function (in durations and/or costs terms), where the project risk level is measured as the variance of the distribution function (Chapman and Ward, 2003; Markowitz, 1959). The literature indicates other project risk measures, such as Value at Risk (Caron et al., 2007; Rezaei et al., 2020) or semi-variance (Zhang et al., 2011). These two measures have been applied mainly to measure risk in finance and to also assess the economic value of the project portfolio.

Some works perform quantitative risk analyses by incorporating stochastic uncertainty together with aleatory uncertainty. For example, Leopoulos et al. (2006) calculate the project risk by adding the exposures of all the identified risks to draw up an efficient schedule and to effectively prepare budgets. Other recent works also resort to Monte Carlo Simulation to calculate the total project risk to determine time and cost contingencies (Allahi et al., 2017; El-Kholy et al., 2020; Eldosouky et al., 2014; Hoseini et al., 2020a; Kwon and Kang, 2019; Traynor and Mahmoodian, 2019).

Regardless of the methodology, these works only consider uncertainty at the beginning of the project (i.e., before the project starts) to estimate the project risk, but the uncertainty of the project changes while it is underway. The first project stages imply the highest uncertainty level because most activities (with their inherent uncertainty) have not yet been performed. However, as project execution advances, the uncertainty level drops (as activities lose their uncertainty once they finish). Accordingly, Pajares and López-Paredes (2011) introduced the Schedule Risk Baseline (SRB) concept to monitor the evolution of the project's aleatory uncertainty while it is underway based on Monte Carlo simulation. Other works into risk management also apply the SRB concept. Acebes et al. (2014) use the SRB concept to determine the optimal project start date in the event of seasonal uncertainty. More recently, Acebes et al. (2020) define indicators based on the SRB to prioritize activities by contemplating their uncertainty.

In this work, we propose an approach that allows project managers to select a schedule with the lowest overall project risk of several schedules with the same duration. To do so, we calculated the overall project risk associated with the aleatory uncertainty of its activities. That is, we analyzed how this aleatory uncertainty (due purely to the random nature of the activity duration) affects the total project risk. Based on previous research, we used the SRB concept to monitor the evolution of uncertainty while the project is underway. As we demonstrate, different schedules for the same project entail a different project risk level. In our approach, we propose employing an indicator (Schedule Risk Value, SRV) to measure the project risk corresponding to different schedules of the same project (all with the same duration). The aim of our contribution is to provide a tool that allows comparisons of the project risk of different schedules (with the same duration) for the same project for project managers to select the schedule with the lowest project risk (i.e., the lowest SRV).

## 3. Integrating aleatory uncertainty into project scheduling

The general procedure to achieve a schedule that meets planning, resources, financing, and risk requirements is summarized in the diagram shown in Fig. 1.

During the Scope Management process, the project, product limits, and acceptance criteria are described in detail (Project Management Institute, 2017). This process results in the definition of the activities to be carried out, which include all the work foreseen in the project. Once the duration of each activity has been estimated and their precedence relationships have been analyzed, a first project schedule can be obtained, normally using the Critical Path methodology that provides a project schedule with the shortest possible duration.

This initial schedule is only viable if all the resources required for performing activities are available on the dates they are scheduled. However, resources are normally scarce, and it is not possible to perform all the activities according to the initial schedule (i.e. the schedule with the shortest duration). Considering resource constraints involves such a high degree of combinatorial complexity that the optimal project schedule is unknown. Consequently, the different methods applied for resource-leveling found in the literature normally provide different schedules for the same project, where the solutions with the shortest duration are normally considered the best-known schedule for a project. Each alternative schedule has a different distribution of activities over time, although precedence relationships are maintained.

After obtaining several alternative schedules that meet that resource constraint, project managers must check whether these schedules have the actual funds available for project execution (Villafáñez et al., 2020). For each schedule, we must check if project funding during each period is enough to cover the expenses incurred in the project. Only those schedules that meet the funding constraint will be financially viable. To this end, one option is to seek new funding alternatives that meet project needs. Another option is to delay performing some project activities until enough funds become available to execute them which, consequently, involves determining a new project schedule.

In this paper, we propose additional steps in the decision-making process that entail selecting a project schedule baseline. These steps begin with the "Risk Management" block (Fig. 1.). Our starting point is different schedules for the same project (with the same duration, cost and resource use) that meet scope, resources and funding constraints. At this point, our objective is to select the schedule with the lowest total risk associated with the aleatory uncertainty in activities' duration. Of the four uncertainty types (aleatory, stochastic, epistemic, ontological, (Hillson, 2014a)), in this article we deal with only aleatory uncertainty, which is the type of uncertainty that is embedded practically in each activity (e.g. range of an activity's duration for many reasons). That is, we analyze how uncertainty in activities' duration (due purely to the random nature of the activity duration) affects the total project risk. To do this, we first add aleatory uncertainty to the duration of the project activities. Then, to select a more adequate schedule, we calculate the total risk associated with all the available alternative schedules.

As the activities in each schedule have different starting and ending dates, the uncertainty introduced by each activity will have a different impact on the calculation of the total project risk depending on the dates it is scheduled. Consequently, different schedules will result in a different project risk level. To assess the total risk associated with each schedule, we built the Schedule Risk Baseline, SRB (Pajares and López-Paredes, 2011), for all the alternative schedules (Subsection 3.1). Once the project risk is evaluated for several execution alternatives (i.e. the SRB values for different schedules), we calculate the Schedule Risk Value, SRV (Acebes et al., 2020), to compare the risk in those possible alternative schedules (Subsection 3.2). The schedules with a lower SRV value have less probability of deviation from their final duration than those with a higher SRV value. Consequently, the schedules with a lower SRV value are more likely to meet the deadline.

### 3.1. Calculating the project risk associated with each alternative schedule.

A baseline is a dataset that serves as a reference for the successive comparisons made of the actual and initial situations of any event.

The risk baseline represents the evolution of the project risk value throughout its life cycle. In this work, we use the Schedule Risk Baseline (SRB) concept introduced by Pajares and López-Paredes (2011) to calculate the risk associated with different schedules of the same project. The procedure to obtain SRB assumes that the duration of project activities follows a probability distribution function. By Monte Carlo simulation, project variance is calculated not only at the beginning, but also at different control points, along with the project schedule. This makes it possible to consider that, at a given control point, the contribution to the total project risk of the activities that have already been completed at that time is zero.

Based on this risk baseline concept, this paper considers that project activities imply some uncertainty in their duration (i.e. aleatory uncertainty according to Hillson, (2014c)). We take one of the feasible project schedules and then apply Monte Carlo simulation to each execution period, from the initial time point, when the project has not yet started, to the final time point, when all the activities have been performed.

The project risk at each time instant corresponds to the variance of the total duration distribution function by considering that the project is executed according to its initial planning. At the initial time point, the project has not yet started and the duration and uncertainty of all the activities remain. Project risk is maximum at this point. At each intermediate control time, some activities will have finished (completely or partly). These activities eliminate the corresponding uncertainty and, by performing a new Monte Carlo simulation, we obtain a new value for the risk corresponding to the uncertainty of the activities that have not yet ended at that control time (i.e. ongoing and unstarted activities). This value is calculated as the variance of the resulting distribution function in this new situation. If we repeat this operation at each control time point, we obtain the project risk value (variance) during each period from the beginning of the project to its end. By joining the consecutive points, we obtain the schedule risk baseline, SRB (Fig. 2.).

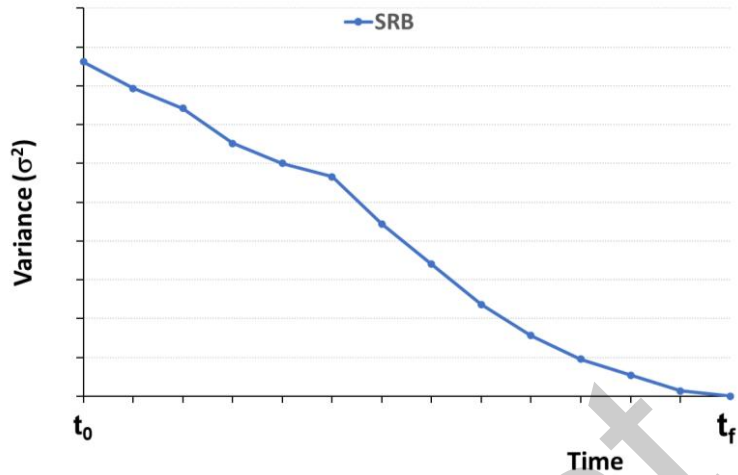

**Fig. 2.** Schedule Risk Baseline (SRB) corresponding to one of the feasible project schedules.

When the project ends, uncertainty disappears because there are no more activities to be performed. Consequently, the value of variance at the end of the project will always be zero (control period $t=t_f$ in Fig. 2.).

We propose applying this procedure to obtain an SRB curve for each feasible schedule (i.e. those alternative schedules with the same duration, cost and resource use) as the first step to calculate the risk associated with each execution alternative. An activity (which introduces a certain level of uncertainty into the project) scheduled at an earlier or later date (provided time, cost, and resource constraints hold) results in a different SRB curve. Consequently, distinct schedules result in differing SRB curves. For example, if the activities that introduce the most uncertainty are scheduled during the initial project periods, their associated uncertainty is soon eliminated, which makes the SRB graph rapidly decrease. However, if these activities are scheduled in a later project stage, the uncertainty introduced by these activities remains while the project is underway, until the activity finally ends and the associated uncertainty disappears. Consequently, the SRB curve continues to display high values until this activity finishes.

### 3.2. Comparing the risk associated with different alternative schedules

As described above, the SRB curve represents the project risk for each execution time corresponding to a particular schedule for the same project. We use this information to calculate the project risk associated with the different feasible schedules for the same project. To this end, we apply the Schedule Risk Value (SRV) concept introduced by Acebes et al. (2020), which is defined as the area under the Schedule Risk Baseline (SRB) curve from the beginning of the project (t=0) to its end (t=$t_f$). This area may be calculated with Eq. 1:

$$SRV = \int_{t=0}^{t=t_f} SRB_t \qquad\qquad \text{Eq. 1}$$

Fig. 3. represents the total project risk (dashed area) before project execution starts (t=$t_0$).

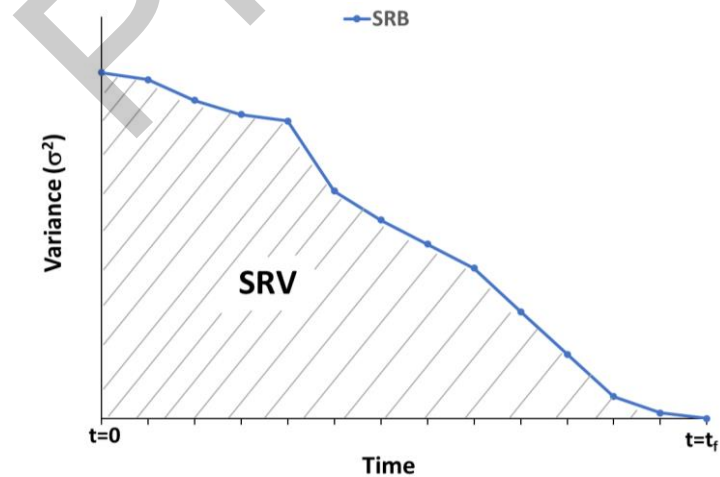

**Fig. 3.** Schedule Risk Value (SRV): area under the SRB curve

Two different schedules for the same project can have the same level of risk at the beginning of the project (i.e. the initial value of the variance of the SRB curve: $SRB1_{t=0} = SRB2_{t=0}$). However, the evolution of their risk while the project is underway and following different schedules can differ. This leads to different SRB curves for distinct project schedules, as shown in Figure 4. This implies that the SRV value (i.e. total project risk) differs for each project schedule.

The fact that the area under the curve of SRB for Schedule 1 is larger than the area under the curve of SRB for Schedule 2 means that the total risk of Schedule 2 is higher than the risk of Schedule 1. As the SRV value depends on the variance of the duration of project activities, a correlation appears between the SRV value and the possibility of finishing the project on time: the higher the SRV value, the greater the uncertainty while the project is underway and, thus, the more chances of not meeting the project end date.

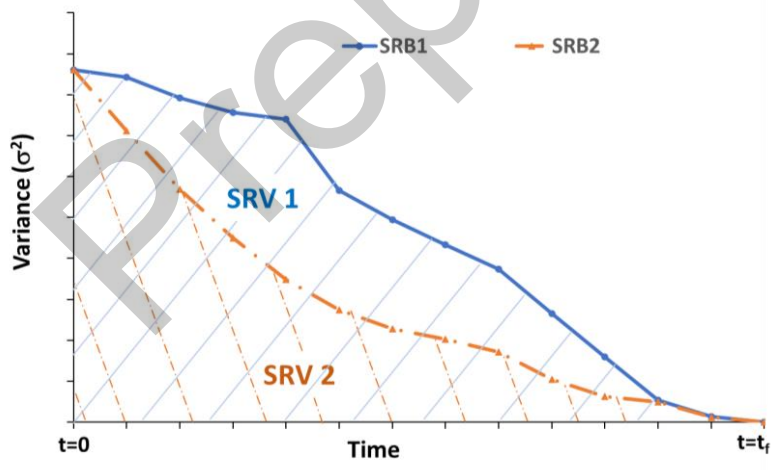

**Fig. 4.** Schedule Risk Value (SRV) for two different schedules for the same project: SRV1 and SRV2

The SRB value for Schedule 2 (Fig. 4) rapidly lowers during the initial project execution periods. This means that the activities that confer the project schedule the most uncertainty are performed during the first project periods, whereas the activities scheduled during the latter project periods only confer marginal uncertainty. Consequently, the probability of the actual project duration coming close to the planned project duration is high.

In contrast, Schedule 1's uncertainty (Fig. 4) does not noticeably drop until its intermediate schedule dates. The representation of SRB shows that uncertainty remains constant, which implies that the activities scheduled for the first project periods do not confer the project much uncertainty. Therefore, during the first project period, the uncertainty in the estimated completion date is high.

This means that different (feasible) schedules for the same project lead to distinct SRB curves and, thus, differing SRV values, which is the indicator that we propose using to compare the risk associated with distinct project schedules. As our starting point is the different schedules for the same project with the same duration, cost and resource use, the calculation and comparison of SRV for all these schedules allows project managers to select the schedule with the lowest total risk.

## 4. Illustration of the method

In this section, we explain how the proposed approach can be applied to select the schedule with the lowest risk from among all the feasible options that project managers have at this point. We apply the SRV calculation to all the execution alternatives to help project managers to make the best decision regarding project risk. The starting point of our approach is a set of different schedules for the same project of the same duration. We also apply the proposed procedure to a practical example using the MPSPLib project library (Homberger, 2007) because it provides a set of different schedules for the same project. This library has a collection of problems with which to simulate and provide different results obtained with several scheduling algorithms used by the scientific community (Wauters et al., 2015). The MPSPLib library is, in fact, a frequently used repository by researchers in an attempt to find different feasible schedules for the same project (Li et al., 2021).

The MPSPLib library is a public library that contains a collection of 140 artificial multiproject problems. These 140 multiproject problems form a combination of several single-project problems previously collected in the PSPLib (Library for Project Scheduling Problems) according to Kolisch and Sprecher (1996). Each problem is made up of a different number of projects (2, 5, 10, or 20) and all these projects can imply a different number of activities (30, 90, or 120). Each problem is a combination of several projects with a similar number of activities, where the availability of the limited resources in each project prevents the activities from being performed according to the initial schedule. Researchers apply their algorithms and upload their proposed solutions. For each problem, all the uploaded solutions are ranked according to several quality indicators of the schedule. This ranking allows the scientific community to compare different schedules obtained by several algorithms for the same problem.

In our case, we focus on the problems comprising two projects (Fig. 5.). The projects in each problem have 30 activities of deterministic durations with different resource constraints (they correspond to instances ID=6 to ID=10).

| INSTANCES | NO. OF JOBS: 30 ▼ | NO OF PROJECTS 2 ▼ | GLOB.RES.: all ▼ | | BEST SOLUTIONS | | | PERFORMANCE CRITERION: APD ▼ | | | METHOD: all methods ▼ | | | |
|---|---|---|---|---|---|---|---|---|---|---|---|---|---|---|
| ID | INSTANCE | NO. OF JOBS/PROJ. | NO. OF PROJ. | NO. OF GLOB.RES. | OLF | NO. OF SOLUTIONS | ORIGINATOR | DATE | APD | TMS | DPD | METHOD | EF | CT | EXP |
| 6 | mp_j30_a2_nr1 | 30 | 2 | 2 | HIGH (1.17) | 78 ⊡ | Gómez/Fernández | 2019-08-02 | 4.5 | 61 | 0.707107 | ACO+SMT | D | 100 | 800 | n/a |
| 7 | mp_j30_a2_nr2 | 30 | 2 | 1 | LOW (0.85) | 63 ⊡ | Tony Wauters | 2012-06-27 | 15 | 60 | 11.3137 | HYPER | C | 100000 | 0 | 14 |
| 8 | mp_j30_a2_nr3 | 30 | 2 | 2 | LOW (0.59) | 60 ⊡ | Trautmann/Homberger | 2008-05-14 | 3 | 65 | 4.24264 | MAS/PS | D | 0 | 0 | n/a |
| 9 | mp_j30_a2_nr4 | 30 | 2 | 3 | LOW (0.79) | 39 ⊡ | Dietz/Homberger | 2008-05-20 | 10.5 | 54 | 7.77817 | CMAS/ES-STV | D | 0 | 0 | n/a |
| 10 | mp_j30_a2_nr5 | 30 | 2 | 1 | LOW (0.77) | 36 ⊡ | Tony Wauters | 2012-06-27 | 8.5 | 58 | 12.0208 | HYPER | C | 10 | 1 | 14 |

**Fig. 5.** Problem instances composed of two projects with 30 activities (jobs) each

For each selected problem, MPSPLib records different solutions, which we sort according to Total Makespan (TMS) and observe several execution alternatives for the same problem with the same TMS. That is, for the same problem (the same initial schedule), we find different feasible schedules that meet the resource constraints with the same duration (same TMS) (Fig. 6.).

**Fig. 6.** Different solutions (schedules) with a duration (TMS) of 65-time units for problem ID=8

All these schedules implement the defined project scope and comply with resource constraints. If we assume that all the solutions are viable in funding availability terms, then any of these schedules can be selected as the schedule baseline because they all have the same duration (TMS = 65 time units). How can project managers select the most adequate schedule in project risk terms? That is to say, among all the known viable schedules with a duration of 65 time units, which schedule offers the highest probability of finishing the project on time?

At this point, we use the decision variable SRV. Of all the possible schedules that have passed all the previous filters, we perform a risk analysis to select from among all the possible schedules with the same TMS that with the lowest total risk (i.e. the lowest SRV indicator value).

The SRV indicator provides information on the certainty of completing the project on the indicated date (65 time units in the present example, problem ID=8) by considering the uncertainty of each activity from the time the project starts to its end. In this case, 13 different schedules correspond to the best-known solution: 65 time units (Fig. 6.).

To obtain the risk (SRV) corresponding to each schedule, we must previously calculate the schedule risk baseline (SRB) by Monte Carlo simulation. To do so, we need to incorporate aleatory uncertainty into the duration of the activities because the value considered for the duration of the activities in the MPSPLib library is deterministic.

Traditionally, different types of distribution functions have been used to generate stochastic durations for activities (Uniform, Beta, Normal, Triangular, etc.) (Vanhoucke, 2012, 2011, 2010). In this paper, we use a lognormal distribution function due to its capability to model variability in the duration of activities (Colin and Vanhoucke, 2016; Trietsch et al., 2012). The values generated by this distribution function type are sufficiently far away from the mean distribution value and do not generate negative values for the duration of activities. Other authors utilize other types of distribution functions to assign uncertainty to activities. For example, Leopoulos et al. (2006) resort to triangular distribution functions; Mohamed et al. (2020) use uniform distribution functions; Acebes et al. (2014a) employ normal distribution functions, while Allahi et al. (2017) and Hoseini et al. (2020b) apply beta distribution functions.

Notwithstanding, the introduced uncertainty type is not relevant for implementing the method that we herein propose because the procedure would be the same as that we describe in this paper. Furthermore in a practical case, project managers can assign the uncertainty type that they believe best fits the activity's stochastic nature (i.e. triangular, beta, uniform, lognormal, or others).

The lognormal distribution function that we apply to model the aleatory uncertainty of activities uses the expected value and the standard deviation as input. The expected values of the duration of activities are those indicated for each schedule in the MPSPLib library. The duration variability is modeled using the coefficient of variation CV ($CV_i = \frac{\sigma_i}{\mu_i}$) (Ballesteros-Pérez et al., 2020). These values are generated randomly for each activity following a uniform distribution that varies between 0.10 and 0.30 (values close to 0.1 mean little variability and those close to 0.3 represent wide variability).

After obtaining the data that characterize each project's activities (expected duration and variability), we calculate the SRB as explained in Section 3.1 (Fig. 7.). This graph allows us to calculate the value corresponding to the total project risk (SRV) as the area under the SRB curve.

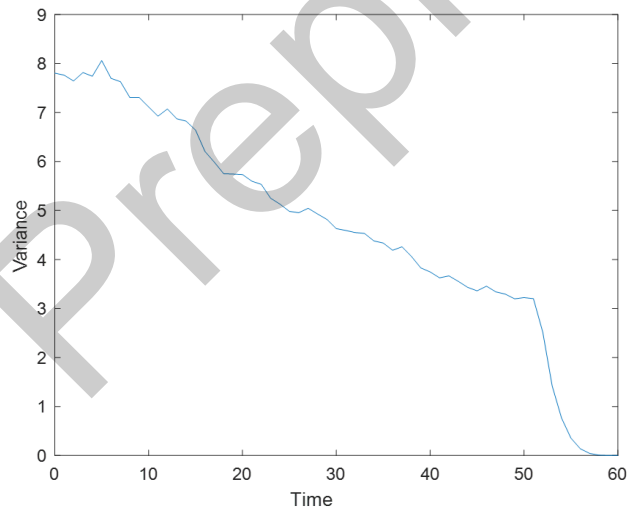

**Fig. 7.** Graphical representation of the SRB for the problem project "mp_j30_a2_nr3"

After calculating the total risk (SRV) for each solution to the problem (i.e. each viable schedule), we select the schedule with the lowest SRV indicator value.

## 5. Results and discussion

In this section we demonstrate that different schedules for the same project (all with the same duration) may entail a distinct risk level. The starting point of our simulation is the eight best solutions for the problem (ID=8) from the MPSPLib library. These eight solutions were provided by different resolution methods, which gave rise to eight differing schedules for the same project (but with the same duration; i.e. Total Makespan, TMS). These possible feasible schedules with the same duration are those which project managers could choose to establish a baseline. In this paper, we propose going one step further and using the level of risk associated with each schedule when establishing this baseline.

Table 1 presents the simulation results corresponding to the eight best solutions to the problem with ID=8 in the MPSPLib library. For each one, we indicate the name of the resolution method that generates each schedule, the planned duration according to MPSPLib (note that the library considers a deterministic duration for activities) and the average duration by considering the uncertainty of activities and the total risk (SRV).

**Table. 1:** Results obtained after simulating eight different feasible schedules with the same duration for the same project (problem mp_j30_a2_nr3 from MPSPLib)

| | Schedule 1 (MAS/PS) | Schedule 2 (MAS/CI) | Schedule 3 (CMAS/ES-BORDA) | Schedule 4 (GT-MAS) | Schedule 5 (PSGSMINSLK) | Schedule 6 (MATHEUR) | Schedule 7 (WPR_GA) | Schedule 8 (ACO+SMT) |
|---|---|---|---|---|---|---|---|---|
| *Planned Duration* | 65,00 | 65,00 | 65,00 | 65,00 | 65,00 | 65,00 | 65,00 | 65,00 |
| *Average Duration* | 66,40 | 66,76 | 66,72 | 66,32 | 66,40 | 66,35 | 66,31 | 66,46 |
| *SRV* | 372,07 | 337,77 | 372,68 | 405,01 | 394,46 | 399,57 | 399,66 | 356,85 |

The planned duration for the eight schedules was 65 time units according to MPSPLib (activities with a deterministic duration). After introducing aleatory uncertainty into the duration of activities and performing the simulation, we observe that the average duration slightly varies between all eight schedules, from a minimum value of 66.31 time units with scheduling method WPR_GA to 66.76 time units with the scheduling algorithm MAS/CI. (Fig. 8.)

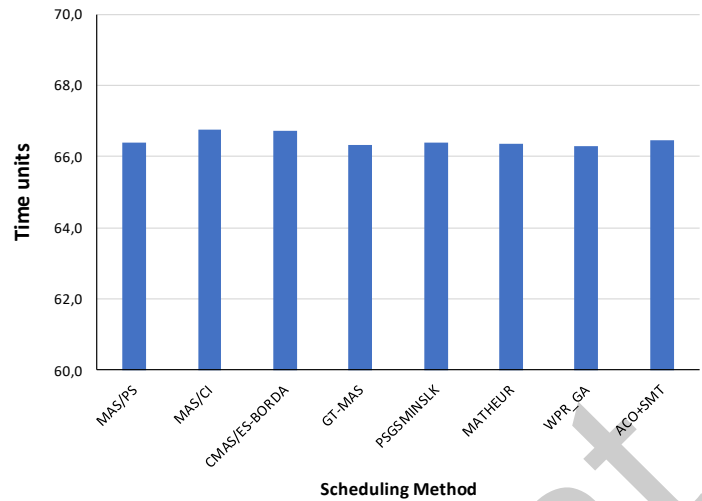

**Fig. 8.** Average duration of the solutions corresponding to each scheduling method

In Fig. 9 we show the risk (SRV) obtained for the eight schedules. We observe that the schedule achieved by method MAS/CI has the lowest risk value, even though its average duration is slightly longer than in the other schedules. This means that the uncertainty of executing the project according to this schedule is less than the uncertainty of the other seven schedules for the same project. The selection of this schedule implies that the deviation from the expected project finish date is more limited than in the other alternative schedules.

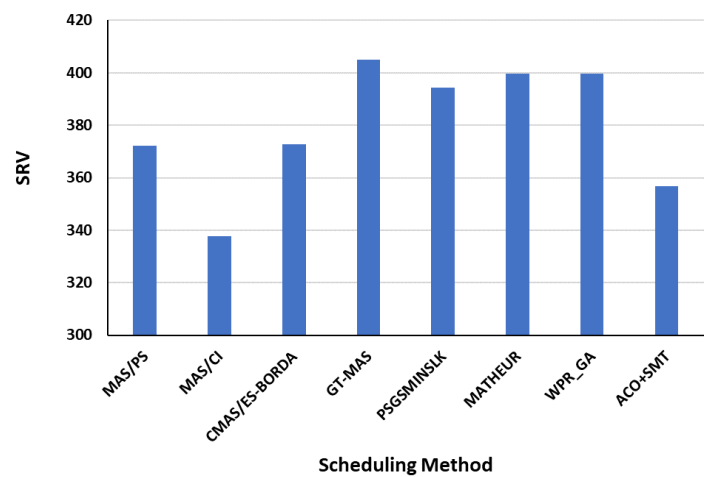

**Fig. 9.** Risk (SRV) associated with the eight feasible schedules for the same project (i.e., solutions with the same duration corresponding to different scheduling methods).

We now extend our analysis to other problems from the MPSPLib library. Table 2 provides the results of the simulations performed with the schedules with the lowest TMS corresponding to the problems with IDs 6 to 10 (Fig. 5). The graphs in this table display the average duration and the value of the indicator SRV corresponding to the schedules obtained by the resolution methods that yielded a schedule with the minimum TMS per problem. Note that these solutions correspond to different schedules (with the same duration) for the same project. The column corresponding to the indicator SRV acts as the basis for selecting the schedule with the lowest total risk.

**Table 2:** Average duration and total project risk (SRV) charts for problems ID 6 to 10 in MPSPLib

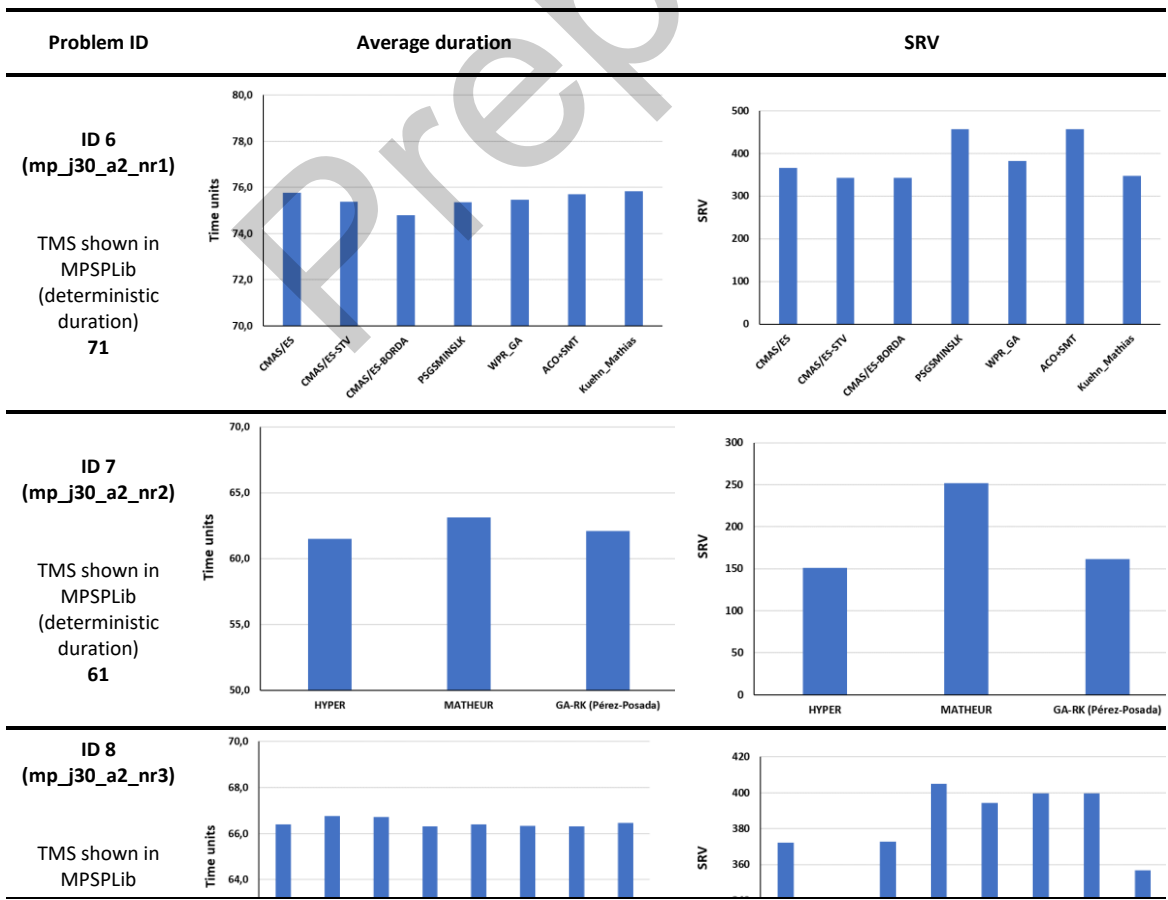

| **ID 9**<br>**(mp_j30_a2_nr4)**<br><br>TMS shown in<br>MPSPLib<br>(deterministic<br>duration)<br>**58** | 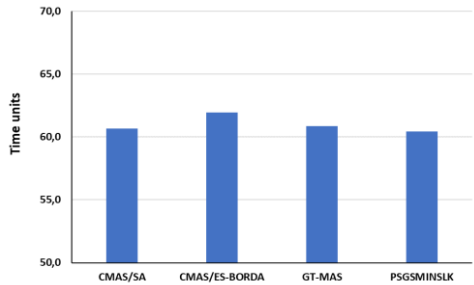 | 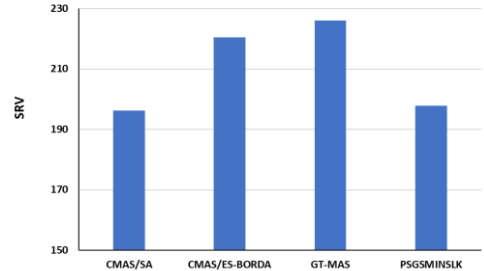 |
| --- | --- | --- |
| **ID 10**<br>**(mp_j30_a2_nr5)**<br><br>TMS shown in<br>MPSPLib<br>(deterministic<br>duration)<br>**58** | 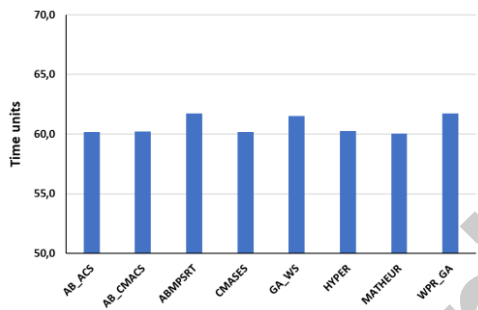 | 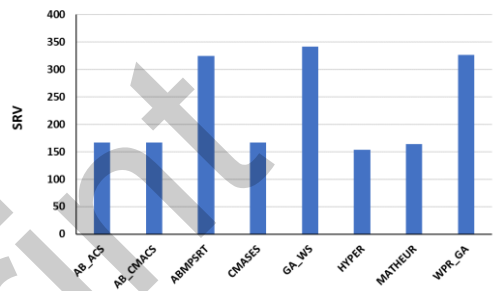 |

From problem "mp_j30_a2_nr1" (ID 6), we observe that the average duration obtained for all the schedules is similar, and the schedule proposed by method CMAS/EN-BORDER is that with the shortest average duration. Regarding the SRV value, this is also the schedule with the lowest risk. Consequently, selecting this schedule as the project baseline is probably a good option. The schedule yielded by method PSGSMINSLK is probably not as advisable in uncertainty terms and, despite having a similar average duration to the other alternatives, our analysis shows that it entails a higher risk.

The results of simulating the schedules for problem "mp_j30_a2_nr2" (ID 7) clearly suggest selecting the schedule provided by method CMAS/SA because it presents the lowest risk (SRV) and also the shortest average duration of all the alternatives.

In problem "mp_j30_a2_nr3" (ID 8), we can see the benefits of using the SRV indicator as a decision tool. The schedule provided by method MAS/CI presents the lowest SRV value. If we observe the average durations of the other alternatives, we find that they all have similar values.

In problem "mp_j30_a2_nr4" (ID 9), the decision to be made also seems clear because two schedules present a lower SRV than the other two alternatives. In this case, the schedule provided by method PSGSMINSLK has the shortest average duration but does not stand out from the other solutions.

Finally for problem "mp_j30_a2_nr5" (ID 10), several solutions appear with a relatively low SRV, and these schedules also have similar average durations. Consequently, we should select the schedule provided by the method with the lowest risk, i.e., the schedule yielded by the HYPER resolution method.

In all the analyzed problems, we use the SRV indicator to determine which viable schedule has the lowest risk value. By doing so, the method herein proposed allows project managers to select the schedule with the highest certainty on the finish date among the alternatives provided by any scheduling method found in the literature.

## 6. Conclusions and future works

From the initial project conception to the approved schedule baseline, a process to select many alternatives is followed to run the project. At some point in the planning phase, some of these schedules are ruled out if they do not meet any project objectives or they do not fulfill scheduling constraints.

The literature on this topic has focused mainly on methods that provide schedules to meet resource restrictions and, more recently, financial limitations. These constraints involve such a high level of combinatorial complexity that the optimal project schedule remains unknown. Consequently, the application of the methods found in the literature provides different schedules for the same project, and the solutions with the shortest duration are considered the best-known schedule for a project. Which of these schedules should become the schedule baseline? To the best of our knowledge, no tools currently allow project managers to select the schedule that can better in project risk terms.

This paper attempts to bridge this gap by presenting a method to allow project managers to select the project schedule with the highest probability of meeting the deadline from several alternative schedules with the same duration. To this end, we integrate project risk into project scheduling by quantifying the risk associated with all the possible known project execution alternatives and the same duration. For this purpose, we employ the SRB/SRV concepts to compare the risk level of several schedules with the same duration. Our approach allows project managers to select the project schedule with the lowest risk (i.e. lowest SRV), and thus the schedule with the highest probability of meeting the deadline from among all the other execution alternatives with the same duration.

We illustrate the usability of these indicators by running a simulation exercise with different projects belonging to the MPSPLib library, a well-known repository for schedule benchmarking. The starting point is the different schedules (with the same duration) for the same project. As the MPSPLib library does not consider activities' uncertainty, we introduced aleatory uncertainty for each activity following the steps included in other related research works. By incorporating uncertainty in the activities, we can calculate the total risk (SRV) associated with each feasible schedule and choose the one with the lowest total risk (and still the same duration). This study used a lognormal distribution function to model activities' aleatory uncertainty. It should be noted that the introduced uncertainty type is not relevant for implementing the method as the procedure is the same as that described in this paper. In a practical case, project managers can assign the uncertainty type that they believe best fits the activity's stochastic nature (i.e., triangular, beta, uniform, lognormal, or others). By taking the best-known schedules (in shortest duration terms) for a project, we demonstrate that the method herein proposed can be used as a decision tool to permit project managers to make better decisions about selecting project schedules with a lower risk from among several alternatives with the same duration.

Among the different types of uncertainty that can impact the total project risk (i.e. aleatory, epistemic, stochastic and ontological), this paper focuses only on aleatory uncertainty. That is, we analyze how this aleatory uncertainty (due purely to the random nature of the activity duration) affects the total project risk. Future research into the integration of project risk into project scheduling can continue with the analysis of the impact of the other three uncertainty types on the total project risk.

Similar studies on assessing risks in portfolio management can be performed by considering the risk associated with the order of executing several projects in a company's portfolio. This study centers on the risk associated with several execution alternatives with the same duration for the same project. Along the same line, a normalized metric can be proposed to compare the uncertainty of projects with different durations in order to assist the decision-making process as to what projects are to be incorporated in a firm's portfolio. Thanks to the proposed method's flexibility, it can be easily adapted to be used in other application domains where decision makers face several alternatives that involve risk and scheduling decisions.

**Acknowledgments**

The authors are grateful to the anonymous reviewers for carefully reading our manuscript, and their many insightful comments and suggestions. The usual disclaimer applies.

## 7. References

Acebes, F., Pajares, J., Galán, J.M., López-Paredes, A., 2014a. A new approach for project control under uncertainty. Going back to the basics. Int. J. Proj. Manag. 32, 423–434.

Acebes, F., Pajares, J., Galán, J.M., López-Paredes, A., 2014b. Exploring the influence of Seasonal Uncertainty in Project Risk Management. Procedia - Soc. Behav. Sci. 119, 329–338.

Acebes, F., Pajares, J., González-Varona, J.M., López-Paredes, A., 2020. Project risk management from the bottom-up: Activity Risk Index. Cent. Eur. J. Oper. Res.

Acebes, F., Pereda, M., Poza, D., Pajares, J., Galán, J.M., 2015. Stochastic earned value analysis using Monte Carlo simulation and statistical learning techniques. Int. J. Proj. Manag. 33, 1597–1609.

Afshar-Nadjafi, B., 2016. A new proactive approach to construct a robust baseline schedule considering quality factor. Int. J. Ind. Syst. Eng. 22, 63–72.

Alavipour, S.M.R., Arditi, D., 2019. Time-cost tradeoff analysis with minimized project financing cost. Autom. Constr. 98, 110–121.

Allahi, F., Cassettari, L., Mosca, M., 2017. Stochastic Risk Analysis and Cost Contingency Allocation Approach for Construction Projects Applying Monte Carlo Simulation, in: World Congress on Engineering. WCE 2017. London.

Alleman, G.B., Coonce, T.J., Price, R.A., 2018. What is Risk? Meas. News 01.

Association for Project Management, 2004. Project Risk Analysis and Management (PRAM) Guide, 2nd ed. ed. APM, High Wycombe, Bucks UK.

Ballesteros-Pérez, P., Narváez, A.C., Mateo, M.O., Fernández, A.P., Vanhoucke, M.,

2020. Forecasting the Project Duration Average and Standard Deviation from Deterministic Schedule Information. Appl. Sci. 10.

Blazewicz, J., Lenstra, J.K., Rinnooy Kan, A.H.G., 1983. Scheduling subject to resource constraints: classification and complexity. Discret. Appl. Math. 5, 11–24.

Caron, F., Fumagalli, M., Rigamonti, A., 2007. Engineering and contracting projects: A value at risk based approach to portfolio balancing. Int. J. Proj. Manag. 25, 569–578.

Chapman, C.B., 2006. Key points of contention in framing assumptions for risk and uncertainty management. Int. J. Proj. Manag. 24, 303–313.

Chapman, C.B., Ward, S., 2003. Project Risk Management: Processes, Techniques and Insights, 2nd ed. ed. Chichester, New York.

Chapman, C.B., Ward, S., 2000. Estimation and evaluation of uncertainty: a minimalist first pass approach. Int. J. Proj. Manag. 18, 369–383.

Colin, J., Vanhoucke, M., 2016. Empirical Perspective on Activity Durations for Project-Management Simulation Studies. J. Constr. Eng. Manag. 142, 04015047–1.

Cox, L.A., 2008. What's wrong with risk matrices? Risk Anal. 28, 497–512.

Crawford, L., Pollack, J., England, D., 2006. Uncovering the trends in project management : Journal emphases over the last 10 years. Int. J. Proj. Manag. 24, 175–184.

Dasović, B., Galić, M., Klanšek, U., 2020. A survey on integration of optimization and project management tools for sustainable construction scheduling. Sustain. 12.

Deshmukh, G.K., Mukerjee, H.S., Prasad, U.D., 2020. Risk Management in Global CRM IT Projects. Bus. Perspect. Res.

Dey, P.K., Clegg, B., Cheffi, W., 2013. Risk management in enterprise resource planning implementation: A new risk assessment framework. Prod. Plan. Control 24, 1–14.

Doskočil, R., 2015. An Evaluation of Total Project Risk Based on Fuzzy Logic. Bus. Theory Pract. 17, 23–31.

Dowie, J., 1999. Against risk. Risk Decis. Policy 4, 57–73.

El-Kholy, A.M., Tahwia, A.M., Elsayed, M.M., 2020. Prediction of simulated cost contingency for steel reinforcement in building projects: ANN versus regression-based models. Int. J. Constr. Manag. 0, 1–15.

El-Sayegh, S.M., Manjikian, S., Ibrahim, A., Abouelyousr, A., Jabbour, R., 2018. Risk identification and assessment in sustainable construction projects in the UAE. Int. J. Constr. Manag. 0, 1–10.

Elazouni, A.M., Gab-Allah, A.A., 2004. Finance-based scheduling of construction projects using integer programming. J. Constr. Eng. Manag.

Eldosouky, I.A., Ibrahim, A.H., Mohammed, H.E.D., 2014. Management of construction cost contingency covering upside and downside risks. Alexandria Eng. J. 53, 863–881.

Elms, D.G., 2004. Structural safety: Issues and progress. Prog. Struct. Eng. Mater. 6, 116–126.

Emblemsvåg, J., Kjølstad, L.E., 2006. Qualitative risk analysis: Some problems and remedies. Manag. Decis. 44, 395–408.

European Commission, 2018. Project Management Methodology. Guide 3.0, Conference Record - IEEE Machine Tools Industry Conference. Publications

Office of the European Union, Brussels / Luxembourg,.

Farooq, M.U., Thaheem, M.J., Arshad, H., 2018. Improving the risk quantification under behavioural tendencies: A tale of construction projects. Int. J. Proj. Manag. 36, 414–428.

Fendley, L.G.G., 1968. Towards the development of a complete multi-project scheduling system. J. Ind. Eng. 19, 505–515.

Fergany, M., El-Nawawy, O., Badawy, M., 2020. Estimation of the Overall risk in Residential Building in Egypt. Int. J. Sci. Eng. Res. 11, 1568–1574.

Frank, M., 1999. Treatment of uncertainties in space nuclear risk assessment with examples from Cassini mission implications. Reliab Eng Syst Safe 66, 203–221.

Gavrysh, O., Melnykova, V., 2019. Project risk management of the construction industry enterprises based on fuzzy set theory. Probl. Perspect. Manag. 17, 203–213.

Ghaffari, M., Sheikhahmadi, F., Safakish, G., 2014. Modeling and risk analysis of virtual project team through project life cycle with fuzzy approach. Comput. Ind. Eng. 72, 98–105.

Hazir, Ö., 2015. A review of analytical models, approaches and decision support tools in project monitoring and control. Int. J. Proj. Manag. 33, 808–815.

Hillson, D., 2014a. How to manage the risks you didn't know you were taking. PMI® Glob. Congr. 1–8.

Hillson, D., 2014b. Managing Overall Project Risk, in: PMI Global Congress Proceedings – Dubai, EAU. pp. 1–9.

Hillson, D., 2014c. How risky is your project – and what are you doing about it ? PMI

Glob. Congr. Proc. - Phoenix, Arizona, USA 1–10.

Hillson, D., 2009. Managing Risk in Projects. Gower Publishing, Ltd.

Hillson, D., 2002a. Defining risk: a debate. J. Inf. Technol. Manag. 15, 11.

Hillson, D., 2002b. Extending the risk process to manage opportunities. Int. J. Proj. Manag. 20, 235–240.

Hillson, D., Simon, P., 2012. Practical Project Risk Management: The ATOM Methodology, Second Edi. ed. Management Concepts Inc, Tysons Corner, Virginia.

Homberger, J., 2007. Multi project scheduling problems [WWW Document]. URL http://www.mpsplib.com/ (accessed 2.15.20).

Hoseini, E., Bosch-Rekveldt, M., Hertogh, M., 2020a. Cost Contingency and Cost Evolvement of Construction Projects in the Preconstruction Phase. J. Constr. Eng. Manag. 146, 05020006.

Hoseini, E., van Veen, P., Bosch-Rekveldt, M., Hertogh, M., 2020b. Cost Performance and Cost Contingency during Project Execution: Comparing Client and Contractor Perspectives. J. Manag. Eng. 36, 05020006.

Howell, D., Windahl, C., Seidel, R., 2010. A project contingency framework based on uncertainty and its consequences. Int. J. Proj. Manag. 28, 256–264.

Hsieh, M.Y., Hsu, Y.C., Lin, C.T., 2018. Risk assessment in new software development projects at the front end: a fuzzy logic approach. J. Ambient Intell. Humaniz. Comput. 9, 295–305.

Hulett, D.T., 2011. Integrated Cost-Schedule Risk Analysis. Gower, Farnham, UK.

International Standards Organisation, 2018. ISO31000:2018 Risk management —

Guidelines. Iso 31000.

Jaafari, A., 2001. Management of risks , uncertainties and opportunities on projects : time for a fundamental shift. Int. J. Proj. Manag. 19, 89–101.

Jordan, G.B., Hage, J., Mote, J., Hepler, B., 2005. Investigating differences among research projects and implications for managers. R&D Manag. 35, 501–512.

Karam, A., Lazarova-Molnar, S., 2013. Recent trends in solving the deterministic resource constrained Project Scheduling Problem. 2013 9th Int. Conf. Innov. Inf. Technol. IIT 2013 124–129.

Kelley, J.E., Walker, M.R., 1959. Critical-Path Planning and Scheduling. Pap. Present. December 1-3, 1959, East. Jt. IRE-AIEE-ACM Comput. Conf. - IRE-AIEE-ACM '59 32, 160–173.

Khedr, M.K., 2006. Project Risk Management Using Monte Carlo Simulation. AACE Int. Trans.

Kimiagari, S., Keivanpour, S., 2019. An interactive risk visualisation tool for large-scale and complex engineering and construction projects under uncertainty and interdependence. Int. J. Prod. Res. 57, 6827–6855.

Kolisch, R., Sprecher, A., 1996. PSPLIB - A Project Scheduling Problem Library.pdf. Eur. J. Oper. Res. 96, 205–216.

Kwak, Y.H., Ingall, L., 2007. Exploring Monte Carlo Simulation Applications for Project Management. Risk Manag. 9, 44–57.

Kwon, H., Kang, C.W., 2019. Improving Project Budget Estimation Accuracy and Precision by Analyzing Reserves for Both Identified and Unidentified Risks. Proj. Manag. J. 50, 86–100.

Leopoulos, V.N., Kirytopoulos, K.A., Malandrakis, C., 2006. Risk management for SMEs: Tools to use and how. Prod. Plan. Control 17, 322–332.

Li, F., Xu, Z., Li, H., 2021. A multi-agent based cooperative approach to decentralized multi-project scheduling and resource allocation. Comput. Ind. Eng. 151, 106961.

Little, T., 2005. Context adaptive agility: managing complexity and uncertainty. IEEE Softw. 22, 28–35.

Liu, G., Zhang, J., Zhang, W., Zhou, X., 2007. Risk assessment of virtual enterprise based on the fuzzy comprehensive evaluation method. IFIP Adv. Inf. Commun. Technol. 251 VOLUME, 58–66.

Malcolm, D.G., Roseboom, J.H., Clark, C.E., Fazar, W., 1959. Application of a technique for research and development program evaluation. Oper. Res. 7, 646–669.

Markowitz, H.M., 1959. Portfolio Selection: Efficient Diversification of Investments.

Millington, D., Stapleton, J., 2005. Developing a RAD standard. IEEE Softw. 12, 54–55.

Mohamed, E., Jafari, P., Abourizk, S., 2020. Fuzzy-based multivariate analysis for input modeling of risk assessment in wind farm projects. Algorithms 13, 1–28.

OGC, 2009. Managing Successful Projects with PRINCE2, 2009th ed.

Pajares, J., López-Paredes, A., 2011. An extension of the EVM analysis for project monitoring: The Cost Control Index and the Schedule Control Index. Int. J. Proj. Manag. 29, 615–621.

Pearson, A.W., 1990. Innovation strategy. Technovation 10, 185–192.

Pellerin, R., Perrier, N., 2019. A review of methods, techniques and tools for project planning and control. Int. J. Prod. Res. 57, 2160–2178.

Pellerin, R., Perrier, N., Berthaut, F., 2020. A survey of hybrid metaheuristics for the resource-constrained project scheduling problem. Eur. J. Oper. Res. 280, 395–416.

Perminova, O., Gustafsson, M., Wikström, K., 2008. Defining uncertainty in projects – a new perspective. Int. J. Proj. Manag. 26, 73–79.

Pinha, D.C., Ahluwalia, R.S., 2019. Flexible resource management and its effect on project cost and duration. J. Ind. Eng. Int. 15, 119–133.

Project Management Institute, 2017. A Guide to the Project Management Body of Knowledge: PMBoK(R) Guide. Sixth Edition. Project Management Institute Inc.

Project Management Institute, 2009. Practice Standard for Project Risk Management. Project Management Institute, Inc., Newtown Square, Pennsylvania 19073-3299 USA.

Ratbe, D., King, W.R., Kim, Y.-G., 1999. The fit between project characteristics and application development methodologies: a contingency approach. J. Comput. Info. Syst 40, 26.

Rezaei, F., Najafi, A.A., Ramezanian, R., 2020. Mean-conditional value at risk model for the stochastic project scheduling problem. Comput. Ind. Eng. 142, 106356.

Rezaie, K., Amalnik, M.S., Gereie, A., Ostadi, B., Shakhseniaee, M., 2007. Using extended Monte Carlo simulation method for the improvement of risk management: Consideration of relationships between uncertainties. Appl. Math. Comput. 190, 1492–1501.

Ruiz-Martin, C., Poza, D., 2015. Project configuration by means of network theory. Int. J. Proj. Manag. 33, 1755–1767.

Schafer, G., 1976. A Mathematical Theory of Evidence, Princeton. ed. Princeton, NY.

Shenhar, A.J., Dvir, D., 2007. Reinventing Project Management: The Diamond
    Approach to Successful Growth and Innovation. Harvard Business School Press,
    Boston.

Taylor, J.C., 2008. Project scheduling and cost control: Planning, monitoring and
    controlling the baseline. J. Ross Publishing.

Traynor, B.A., Mahmoodian, M., 2019. Time and cost contingency management using
    Monte Carlo simulation. Aust. J. Civ. Eng. 17, 11–18.

Trietsch, D., Mazmanyan, L., Gevorgyan, L., Baker, K.R., 2012. Modeling activity
    times by the Parkinson distribution with a lognormal core: Theory and validation.
    Eur. J. Oper. Res. 216, 386–396.

Turner, J.R., Cochrane, R.A., 1993. Goals and methods matrix: coping with projects
    with ill defined goals and/or methods of achieving them. Int. J. Proj. Manag. 11,
    93–102.

Vanhoucke, M., 2012. Project Management with Dynamic Scheduling: Baseline
    Scheduling, Risk Analysis and Project Control. Springer.

Vanhoucke, M., 2011. On the dynamic use of project performance and schedule risk
    information during project tracking. Omega 39, 416–426.

Vanhoucke, M., 2010. Measuring Time. Improving Project Performance Using Earned
    Value Management, International Series in Operations Research & Management
    Science: Vol. 136. Springer.

Villafáñez, F.A., Poza, D., López-Paredes, A., Pajares, J., 2018. A unified nomenclature
    for project scheduling problems (RCPSP and RCMPSP). Dir. y Organ. 64, 56–60.

Villafáñez, F.A., Poza, D., López-Paredes, A., Pajares, J., Acebes, F., 2020. Portfolio scheduling: an integrative approach of limited resources and project prioritization. J. Proj. Manag. 5, 103–116.

Villafáñez, F.A., Poza, D., López-Paredes, A., Pajares, J., Olmo, R. del, 2019. A generic heuristic for multi-project scheduling problems with global and local resource constraints (RCMPSP). Soft Comput. 23, 3465–3479.

Vose, D., 2008. Risk Analysis: a Quantitative Guide, 3rd ed. ed. Wiley, Chichester, U.K.

Ward, S.C., 1999. Assessing and managing important risks. Int. J. Proj. Manag. 17, 331–336.

Wauters, T., Verbeeck, K., De Causmaecker, P., Vanden Berghe, G., 2015. A learning-based optimization approach to multi-project scheduling. J. Sched. 18, 61–74.

Williams, T.M., 1995. A classified bibliography of recent research relating to project risk management. Eur. J. Oper. Res. 85, 18–38.

Wirawan, J.A.B., Garniwa, I., 2018. Risk analysis development of solar floating power plant in the sea with Monte Carlo method. Proc. - 2018 3rd Int. Conf. Inf. Technol. Inf. Syst. Electr. Eng. ICITISEE 2018 396–401.

Xie, G., Zhang, J., Lai, K.K., 2006. Risk avoidance in bidding for software projects based on life cycle management theory. Int. J. Proj. Manag. 24, 516–521.

Zhang, W.G., Mei, Q., Lu, Q., Xiao, W.L., 2011. Evaluating methods of investment project and optimizing models of portfolio selection in fuzzy uncertainty. Comput. Ind. Eng. 61, 721–728.